\documentclass{elsart}
\usepackage{latexsym}
\usepackage{amssymb}
\usepackage{epsfig}
\def\nuc#1#2{\relax\ifmmode{}^{#1}{\protect\text{#2}}\else${}^{#1}$#2\fi}
\begin{document}
\renewcommand{\baselinestretch}{1.6}
\begin{frontmatter}
\title{Cosmogenic activation of Germanium and its reduction for low
 background experiments}
\small \author[INR]{I. Barabanov} \author[ITEP,INR]{S.Belogurov}
\footnote{Corresponding
author. Tel.: 7 095 1299419; fax: 7 095 8839601; \\
e-mail: belogurov@itep.ru}, \author[INR]{L. Bezrukov},
\author[INR]{A. Denisov},\author[ITEP,INR]{V. Kornoukhov},
 \author[INR]{N.~Sobolevsky}
\address[INR]{\hbox{INR RAS, pr-t. 60 anniv. of October, 7a, 117312
Moscow, Russia}}
\address[ITEP]{\hbox{SSC ITEP, B.
Cheremushkinskaya 25, 117259 Moscow, Russia}}

\begin{abstract}
Production of $^{60}$Co and $^{68}$Ge from stable isotopes of Germanium 
by nuclear
 active component of cosmic rays is a 
principal background source for a new generation of $^{76}$Ge double beta decay 
experiments like GERDA and Majorana. The biggest amount of cosmogenic activity
 is expected to be  produced during transportation of either  enriched material 
or already grown crystal.

In this letter properties and feasibility of a movable iron shield are 
discussed.
Activation reduction factor of about 10 is predicted by simulations with SHIELD 
code for a simple 
cylindrical configuration. It is sufficient for  GERDA Phase II background 
requirements. Possibility of further increase of reduction factor  
 and physical limitations are considered.  Importance of
 activation reduction during Germanium  purification and detector manufacturing
 is emphasized. 

 \end{abstract}
 \end{frontmatter}
\newpage

\section{Introduction}
Search for neutrinoless double beta decay (DBD) is of fundamental importance 
for Physics.  As follows from oscillation experiments, neutrino is a massive
 particle. If the neutrino mass has Majorana nature, neutrinoless DBD may be 
observed  \cite{elliott}. Development of germanium semiconductor detectors, 
availability of
 enriched material, and sufficiently high transition energy
make $^{76}$Ge one of the best nuclei for neutrinoless DBD search. First
 indication of $^{76}$Ge neutrinoless DBD   is done in \cite{klapdor}. GERDA is a 
new $^{76}$Ge DBD experiment, recently accepted at LNGS \cite{gerda}.  
Another advanced project, Majorana is ready to be started in US \cite{major}. 
Background analysis of both the experiments has shown, that the principal
 background source is the internal activity of Germanium crystals due to 
cosmogenic isotopes of $^{60}$Co and $^{68}$Ge \cite{gerda,major,avi,isp}. 
Without special efforts aimed to reduction of cosmogenic activation,
background index in the energy range near Q$_{\beta\beta}$ is hard to be pushed 
below 0.01 -- 0.02 cpy/keV/kg \cite{gerda} (here cpy stands for counts per 
year). The biggest amount of cosmogenic activity
 is expected to be  produced during transportation of the enriched material to
 crystal growth facility, then to detector manufacturer and finally to the 
laboratory.

In order to achieve the level of total background index of 10$^{-3}$ cpy/kg/keV,
required for the GERDA experiment we propose to use a movable shielding 
container for germanium transportation.
In this paper we report simulation results for a sample configuration of such 
a container.  Possibility of shielding properties optimization
and limitations on feasible reduction factor  for cosmogenic activations
are discussed. \\

\subsection*{Relevant physical processes}
At the Earth surface, formation of radioactive isotopes is caused mostly by
 spallation reactions
 of fast nucleons from cosmic rays. Smaller contributions are due to capture 
of stopped negative muons  and muon induced fast neutrons.

Only one dangerous isotope --- $^{60}$Co may be produced after muon capture. 
However the probability of this channel should be much less than 10$^{-4}$,
see e.g. \cite{wytten}. Muon capture rate at the sea level is 
$\approx 10^{-6}$~1/g/s
\cite{charal}. So $^{60}$Co production rate should be much less than 
$10^{-2}$ 1/kg/day.
Contribution of 
muon induced fast neutrons   may be roughly estimated 
using the result of Cocconi \cite{cocconi}, that only about 2\% of nuclear 
disintegrations
by cosmic rays are due to muon induced neutrons.
Thus about 98\% of cosmogenic activations are produced by nuclear active 
component (N-component) of 
cosmic rays and might be attenuated by movable iron shield.
It is necessary to stress, that muon induced contribution practically does 
not decrease in a shield of about 1000 g/cm$^2$ thickness.  

Composition of N-component of cosmic rays at
 sea level 
is the  following: more than 95\% neutrons about 3\% protons and about 2\% 
pi-mesons 
\cite{hayakava,ziegler1}.
Though production rates of neutrons and protons in  nuclear disintegration
 cascades are close to each other, the flux of protons is generally smaller
 and harder, because protons   are stopped
 more efficiently than neutrons due to electromagnetic losses.

For our study we used the energy spectra and fluxes of neutrons and protons
 from  \cite{ziegler2}, fig.\ref{initsp}. Angular distribution is 
supposed to be proportional to $cos^{3.5}\theta$, where $\theta$ is  
zenith angle.  Analysis of uncertainties of 
the flux and spectral data may be found in \cite{ziegler1}.\\

Understanding of N-component propagation and attenuation
in matter is crucial for an effective shield design.
There are a lot of papers devoted to attenuation of N-component of cosmic rays
in atmosphere, water and some other substances \cite{izgorshkova,cocconi,
ziegler1}.
 Special efforts were
 applied to measurements and simulation
of artificial fast neutrons fluxes attenuation in matter \cite{atten}.

 Typical attenuation lengths for N-component in some 
of materials are shown in the \mbox{table. 1.  \cite{izgorshkova,ziegler1}.}
Note however, that attenuation length might be correctly  defined only for 
equilibrium energy spectrum.

 One can see that
 attenuation
length increases with A. This can be easily understood because
interaction cross sections for reactions in a hadronic cascade are 
roughly proportional to A$^{0.66...0.8}$ \cite{toneev}. It worth mentioning,
 that 
for selection of a material for   movable shielding not only 
attenuation length of N-component, but also the density are important. 
Increase of density leads to decrease of the required mass of a container.

From semi-empirical consideration we have chosen iron as an optimal material 
for  shielding container. \\

Another important entry point for simulation  of cosmogenic background
 reduction is knowledge of partial cross sections for 
production of $^{60}$Co and $^{68}$Ge in spallation reactions.

There exist a few physical models describing the spallation process. 
\cite{nm}. Some of the models together with compilation of experimental data
are implemented in the nucleon transport codes like LAHET, SHIELD, FLUKA etc. 
General analysis of the most diffused codes is done in \cite{nm}, see also
 \cite{isp}. 

\subsection*{ The tool: SHIELD code}

An appropriate nucleon transport code may be used for simulation of both
 hadron propagation and radioactive isotope production. Our choice is the 
SHIELD transport code. 

The SHIELD transport code have been elaborated as an universal tool for study 
of the interaction of high energy particles with a matter. 
 The SHIELD code includes the Russian models of nuclear 
reactions, developed at JINR (Dubna) and INR RAS (Moscow), providing simulation 
of all stages of inelastic nuclear interactions in the exclusive approach. The 
SHIELD code have been benchmarked extensively and showed good agreement 
with experimental results for multiple phenomena.

The modern version of the SHIELD code \cite{nm1,nm2} allows to simulate 
interaction of nucleons, pions, kaons, antinucleons, muons and arbitrary (A,Z) 
nuclei with complex extended targets in energy range up to 1 TeV/nucleon. The 
ionization losses, straggling and multiple Coulomb 
scattering are taken into account.  
Transport of neutrons 
below 14.5 MeV is simulated by means of the original neutron transport code 
LOENT based on a 28 group neutron data system ABBN \cite{Abagyan}.

Capabilities of a hadron transport code depend substantially on the
 generator of inelastic nuclear interactions used. In the SHIELD code the 
MSDM generator (Multy Stage Dynamical Model) \cite{nm3} is used. 
Details about used physical models may be found in
refs.\cite{nm4,nm5,nm6,nm7,nm8,nm9,nm10}

\section{Results and discussion}
\subsection*{Simulation details}
 An iron  container shown in  the
fig.\ref{body} was designed for the transportation of enriched germanium
 for the Phase II of GERDA experiment. Container size is $\phi$140~cm x 126.5~cm 
There is a cavity in the container $\phi$54~cm x 40~cm. The cavity is situated
in such a way that the bottom thickness is 15~cm. 

 Total mass of the container is 14.5 tons. In the central part of the cavity
a germanium cylinder $\phi$42 cm x 27  cm is placed as a target. 
In the simulation model a 3 m thick slab of ground  was included. The container
is placed 1.2 m above ground, as it were in a truck.

Entire configuration consisting of container
and germanium target has been simulated. For understanding of reduction factors
activation of germanium   target without iron container was also simulated.

In order to make possible a comparison of our simulation with literature,
and for giving a key to the shield optimization we  discuss 
in addition two issues:     \\

(1) excitation functions for
$^{60}$Co and $^{68}$Ge production by neutrons and protons 
and                           \\
(2) shielding properties of the container including energy spectra
 of nucleons inside the cavity  \\

\subsection*{Isotope production rates}

 In the tables 2,3 $^{68}$Ge and $^{60}$Co
 production rates  are reported for all the stable isotopes of Ge with and 
without the shielding container. 
Using this table one can predict activation rate and reduction  factor for any 
isotope composition, e.g. 
for enriched $^{76}$Ge (87\% $^{76}$Ge+13\% $^{74}$Ge) reduction factors are 
expected to be  17 for $^{60}$Co and 10 for $^{68}$Ge.
 Taking into account limitation 
due to muon interactions total reduction factor will be about 13
for $^{60}$Co and about 8.5 for $^{68}$Ge.
i.e. sufficient for GERDA phase II experiment.

Attention should be paid to the fact, that although proton flux at the sea
level is only about 3\% of neutron one, its contribution to isotope production
is about 10\% without shielding and up to 20\% with shielding. Contribution of
the sea level protons is not negligible due to hardness of their spectrum
 compare  to  neutron one.

\subsection*{Excitation functions}

Excitation functions for production of  $^{60}$Co and 
$^{68}$Ge by neutrons and protons on stable isotopes of germanium 
were generated with  SHIELD  code.
  Results for  protons coincide within  factor 1.5-2 with experimental
data from \cite{excit,excit1}. 
The most important for our application curves are shown in the fig.\ref{excit}.

Previous estimations of cosmogenic activation of germanium detectors were
done using excitation functions calculated with ISABEL code \cite{avi},
 those results were 2-6 times lower than ours, however such
discrepancy between different methods is typical for this kind of simulation. 
It is important, that accuracy of activation reduction factor of a shield does
 not suffer from these  discrepancies.

\subsection*{Spectra and fluxes of nucleons}

The most important characteristics of a radiation field inside  the cavity 
is differential flux density of nucleons ($J_{\mathrm{N}}(\varepsilon)$).
In the fig.\ref{specinsn} such curves are shown.
The first approximation calculations may take into account
only  neutron spectrum at the sea level and only neutrons inside the cavity.
However,  
one can see, that contribution of the sea level protons  to 
generation of nucleons inside the cavity is more than 15\%
even though their flux is only 3\% of total sea level flux.
This statement is consistent with results of the tables 2,3.

Production rate $R_i$ of ($i$) radioactive isotope may be  found with sufficient 
accuracy
(in our case discrepancy is less than 10\%)
 using the following formula:
$$
R_i=\sum_{\mathrm{\small N}}\sum_jN_j\int J_{\mathrm{\small N}}(\varepsilon) 
\sigma_{ij\mathrm{\small N}}(\varepsilon) d\varepsilon,
$$

where $N_j$ is a number of ($j$) targety nuclei, $\sigma_{ij\mathrm{\small N}}
(\varepsilon) $-
excitation function for ($i$) product at ($j$) target by $\mathrm{\small N}$
 ($\mathrm{\small N}$= n,p) projectile.

It is interesting to compare attenuation length for N-component obtained in our 
simulation with other data. In the fig.\ref{topatten} energy differential
 fluxes of nucleons
 crossing the upper plane of the cavity are shown for two configurations -- with
 and without iron container.  One can see that unique attenuation length can not
 be introduced for both neutron and proton component and for the whole energy
range. It means that spectrum is not an equilibrium one, three main reasons
 could   be pointed out for this fact: systematic errors of determination of 
spectra in  \cite{ziegler2}, systematic errors of simulations 
(including high energy cut), and  different
equilibrium spectra of hadronic shower in air and iron.

One can see that the maximum attenuation length for neutrons is about 
240 $g/cm^2$ at 80 MeV - 200 MeV energy range.  Normally attenuation length is
 measured for neutrons with energy
below 50 MeV. Thus, taking into account energy dependence of attenuation
 length, agreement of our result with table~1  is quite good. 

Most of the activations  are produced by neutrons with energy around 100 MeV,
hence for conservative estimation of shielding properties of an iron container
 attenuation length  of  240 $g/cm^2$ should be used.

Information about spatial distribution of nucleons in the cavity is useful 
for a container shape optimization.  In the fig.\ref{sides} energy differential
 fluxes
 of nucleons    from top, lateral, and bottom surfaces of the cavity  
 are shown.

The most intensive (per unit area) and hardest spectrum 
 comes from the top surface.
Spectrum from bottom is the least intensive and soft. From these curves
the first shape optimization is rather obvious -- iron disk from the bottom
 part can be removed and thickness of the top part may be increased,
 keeping the container mass constant. Simulation of such a modified container
was done. In the fig.\ref{sides} total fluxes into the cavity are shown for the
two configurations. One can observe                                           
$\approx$ 20\%  decrease of nucleon fluxes.
Taking into account muon induced contribution, activation reduction
 for a container with thicker top part are
  10 and 15 for $^{68}$Ge and $^{60}$Co  respectively.

The last configuration will be used for transportation of $^{76}$Ge for 
GERDA experiment.

\subsection*{Prospects}

Further development of $^{76}$Ge DBD  experiments 
will request for few hundred 
kilograms of target isotope and background index better than
 $10^{-3}$ cpy/keV/kg.
The last objective may be reached by combination of sophisticated background
rejection techniques and more efficient shielding against cosmogenic 
activations.

We have shown that significant reduction of activation is possible 
	during transportation. Now the biggest contribution should arrive 
	during crystal growth and detector manufacturing. Construction of 
	stationary shielding above technological equipment should be considered
 as a next step in reduction of cosmogenic background. 

Besides, shape of the container may be optimized within fixed mass.
It is also necessary to  understand better the limitation due to 
 muon-nuclear 	interactions. In particular,  dependence of
secondary fast neutron yield on the material should be investigated.\\

\section{ Conclusions}

Movable iron shielding container is proposed for reduction of cosmogenic 
activations of $^{76}$Ge for DBD experiment. Relevant physical processes are 
considered. Semi-empirical statements useful for a container design 
optimization are 
formulated. Simulation of a simple cylindrical configuration is performed.
Estimation of limitations due to interactions of energetic 
muons is done. Expected reduction factors are  10 and 15 for $^{68}$Ge and 
$^{60}$Co production respectively. The proposed container is built and is
 being used for germanium transportation.

\section*{Acknowledgements}
Authors are grateeful to Prof.  A. Caldwell and Dr. M. Altmann from
MPP, Munich  for stimulating interest. The work is supported by
CRDF grant GAP-1480.

\newpage
\begin{table}
\caption{\mbox{Cosmic Ray neutrons attenuation lengths}}
\label{table_example}
\vspace{3mm}
\centering
\begin{tabular}{|c|c|}
\hline
 Material& $\lambda$, g/cm$^2$\\
\hline
Air& 140-160\\
\hline
concrete& $\approx$170\\
\hline
Iron& $\approx$200\\
\hline
Lead &$\approx$300\\
\hline
\end{tabular}
\end{table}
\clearpage
\newpage
\begin{table}
\caption{$^{68}$Ge production rates(per day, 
per 1 kg), statistical standard deviations are shown in parentheses. }
\label{table_example}
\vspace{3mm}
\centering
\begin{tabular}{|c|c|c|c|c|}
\hline
target&\multicolumn{2}{c|}{total}&\multicolumn{2}{c|}{by sea level protons}\\
\hline
&no shield&shield &no shield&shield\\
\hline
$^{70}$Ge&281.4 (0.5\%)& 33.0 (2\%)&17.17 (1.1\%)&4.90  (1.5\%) \\
\hline
$^{72}$Ge&55.34 (1.4\%)& 6.20 (4\%) &4.78  (2\%)&0.96  (3\%)\\
\hline
$^{73}$Ge&28.0  (1.3\%)& 2.94 (7\%)&2.54  (3\%)& 0.45  (6\%)\\
\hline
$^{74}$Ge&14.53  (2\%)& 1.46 (8\%)&1.48  (4\%)& 0.24  (6\%)\\
\hline
$^{76}$Ge&4.22   (4\%)& 0.4 (8\%)&0.54  (6\%)& 0.06 (12\%)\\
\hline
\end{tabular}
\end{table}
\clearpage
\newpage
\begin{table}
\caption{$^{60}$Co production rates(per day, 
per 1 kg), statistical standard deviations are shown in parentheses.}
\label{table_example}
\vspace{3mm}
\centering
\begin{tabular}{|c|c|c|c|c|}
\hline
target&\multicolumn{2}{c|}{total}&\multicolumn{2}{c|}{by sea level protons}\\
\hline
&no shield&shield &no shield&shield\\
\hline  
$^{70}$Ge&1.73  (7\%)& 0.118 (33\%) &0.170   (11\%)&0.028   (19\%)\\
\hline  
$^{72}$Ge&2.88  (6\%)& 0.256 (19\%) &0.285  (9\%)&0.046   (14\%) \\
\hline  
$^{73}$Ge&3.14  (4.0\%)& 0.265 (24\%) &0.335  (8\%)&0.035   (21\%)\\
\hline  
$^{74}$Ge&3.35  (4\%)& 0.23 (21\%) &0.380  (8\%)&0.050   (14\%)\\
\hline  
$^{76}$Ge&3.31  (4\%)&  0.156 (13\%)&0.455  (7.0\%)&0.036   (15\%)\\
\hline
\end{tabular}
\end{table}

\clearpage
\newpage
 FIGURE CAPTIONS \\

\begin{figure*}[ht]
\caption{Nucleon flux density spectra at the sea level \cite{ziegler2}. 
Asterisks -- neutrons, open triangles -- protons. }
\label{initsp}   
    \end{figure*}

\begin{figure*}[ht]
\caption{Outline of the iron shielding container}
\label{body}   
    \end{figure*}

\begin{figure*}[ht]
\caption{Excitation functions for $^{60}$Co and $^{68}$Ge production by
 neutrons on stable isotopes of Ge.}
\label{excit}   
    \end{figure*}
\begin{figure*}[ht]
\caption{Nucleon flux density spectra inside the cavity. Open triangles -- neutrons,
asterisks -- neutrons from sea level neutrons only, open circles -- protons, black 
triangles -- protons from sea level neutrons only. }
\label{specinsn}   
    \end{figure*}
\begin{figure*}[ht]
\caption{ Flux of nucleons through the top surface of the cavity in two 
configurations -- with and without iron container.
Black squares and circles -- sea level neutrons and protons respectively,
asterisks and black triangles -- neutrons and protons inside the cavity.
 }
\label{topatten}   
    \end{figure*}

\begin{figure*}[ht]
\caption{ Fluxes of nucleons in the cavity:  total and
 from different sides of the cavity.     }
\label{sides}   
    \end{figure*}

\clearpage
\newpage
\setcounter{figure}{0}

\begin{figure*}[ht]
\begin{center}
\epsfig{file=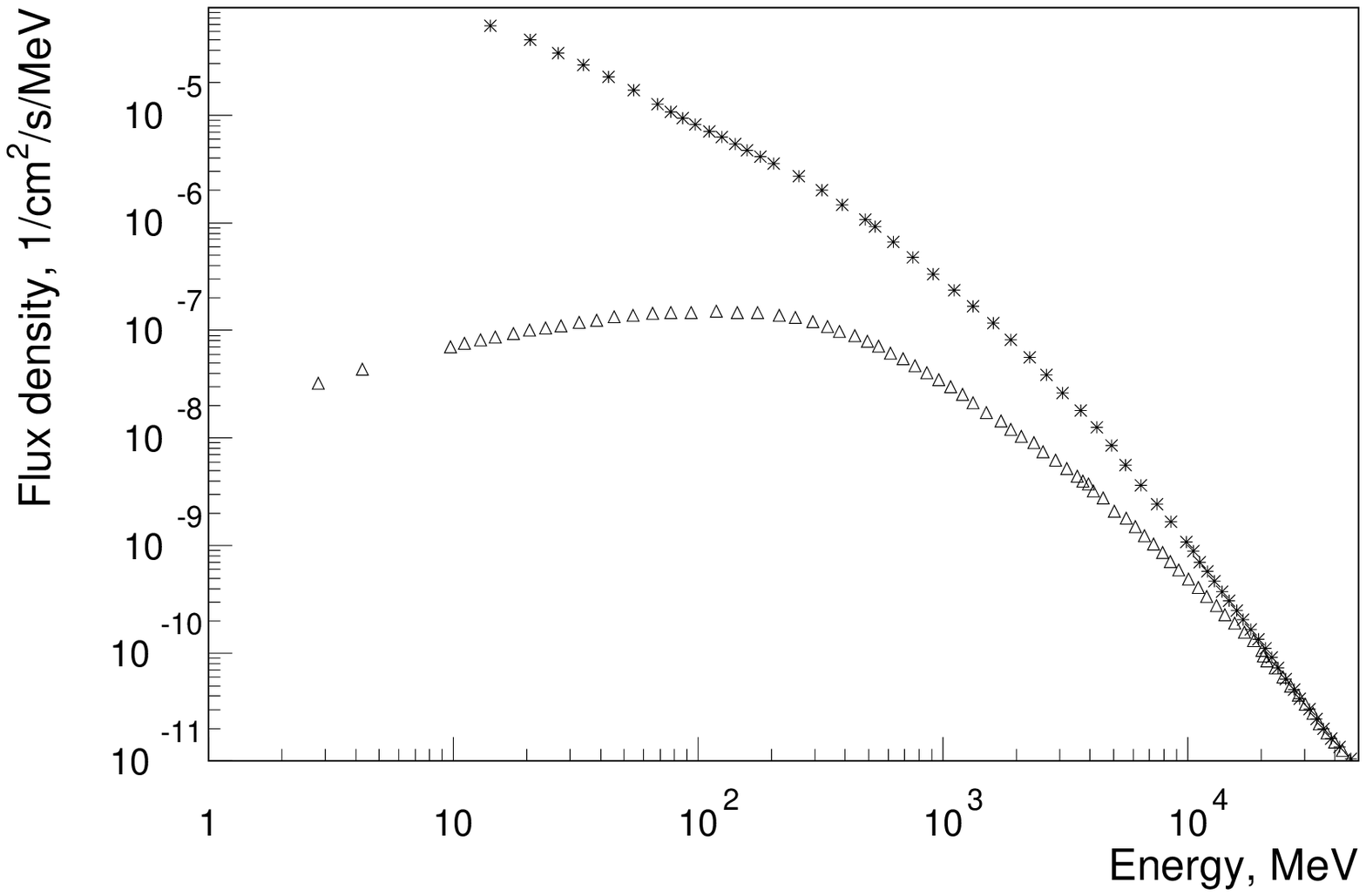,width=16cm}
\end{center}
\caption{}
\label{initsp}   
    \end{figure*}
\newpage
\begin{figure*}[ht]
\begin{center}
\epsfig{figure=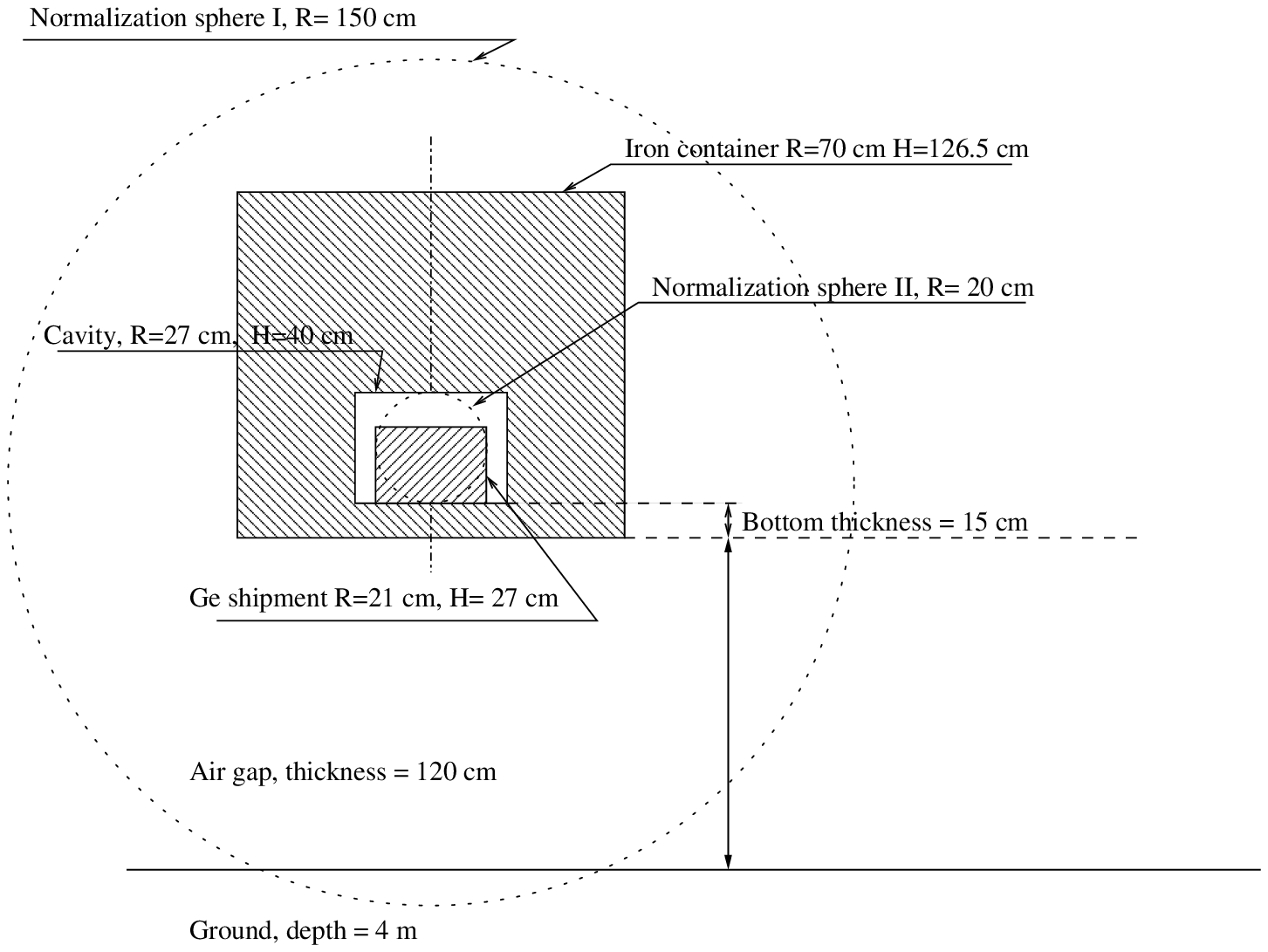,width=16cm}
\end{center}
\caption{}
\label{body}
    \end{figure*}
\newpage
\begin{figure*}[ht]
\begin{center}
\epsfig{figure=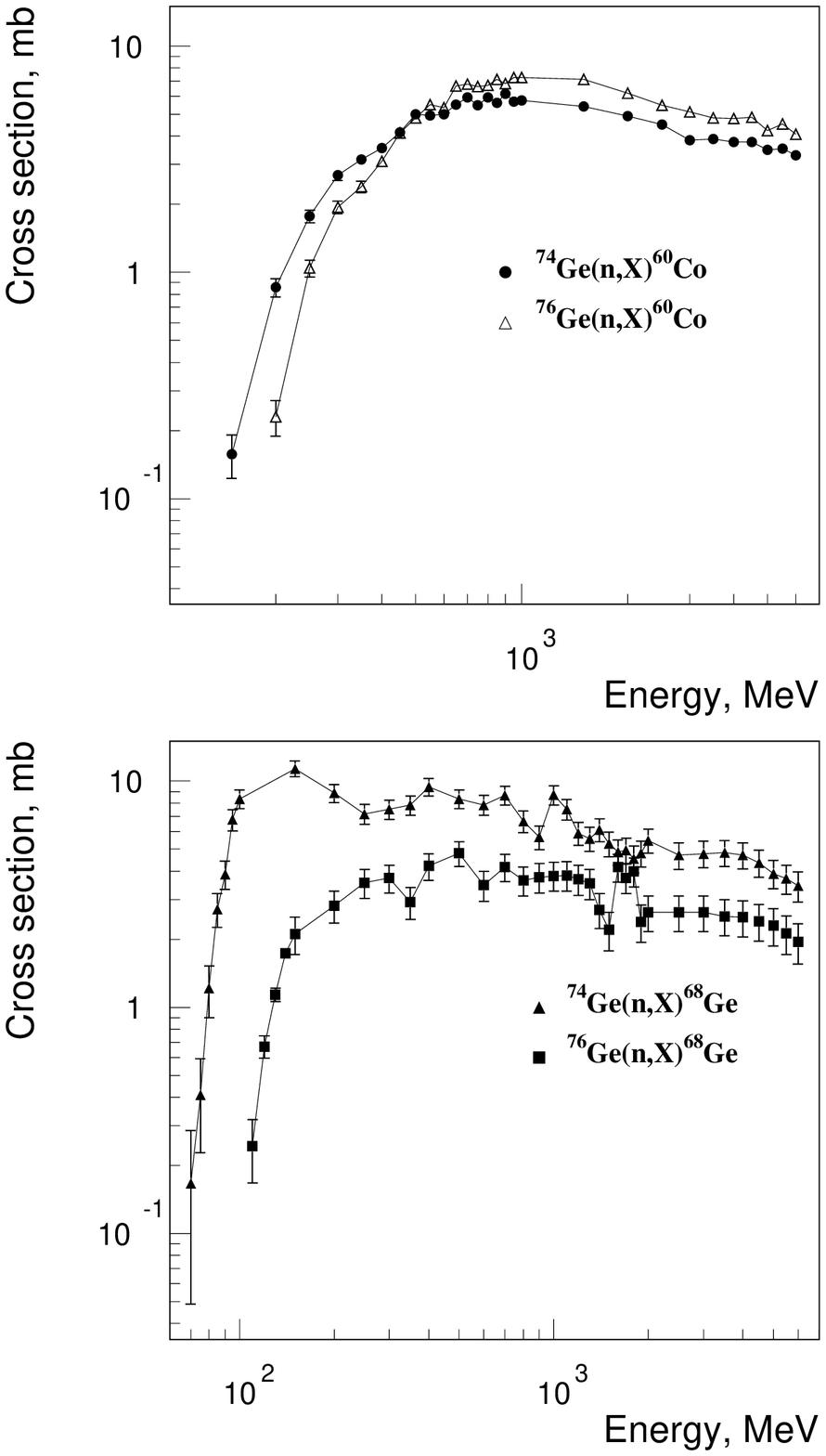,width=13cm}
\end{center}
\caption{}
\label{excit}   
    \end{figure*}
\newpage
\begin{figure*}[ht]
\begin{center}
\epsfig{figure=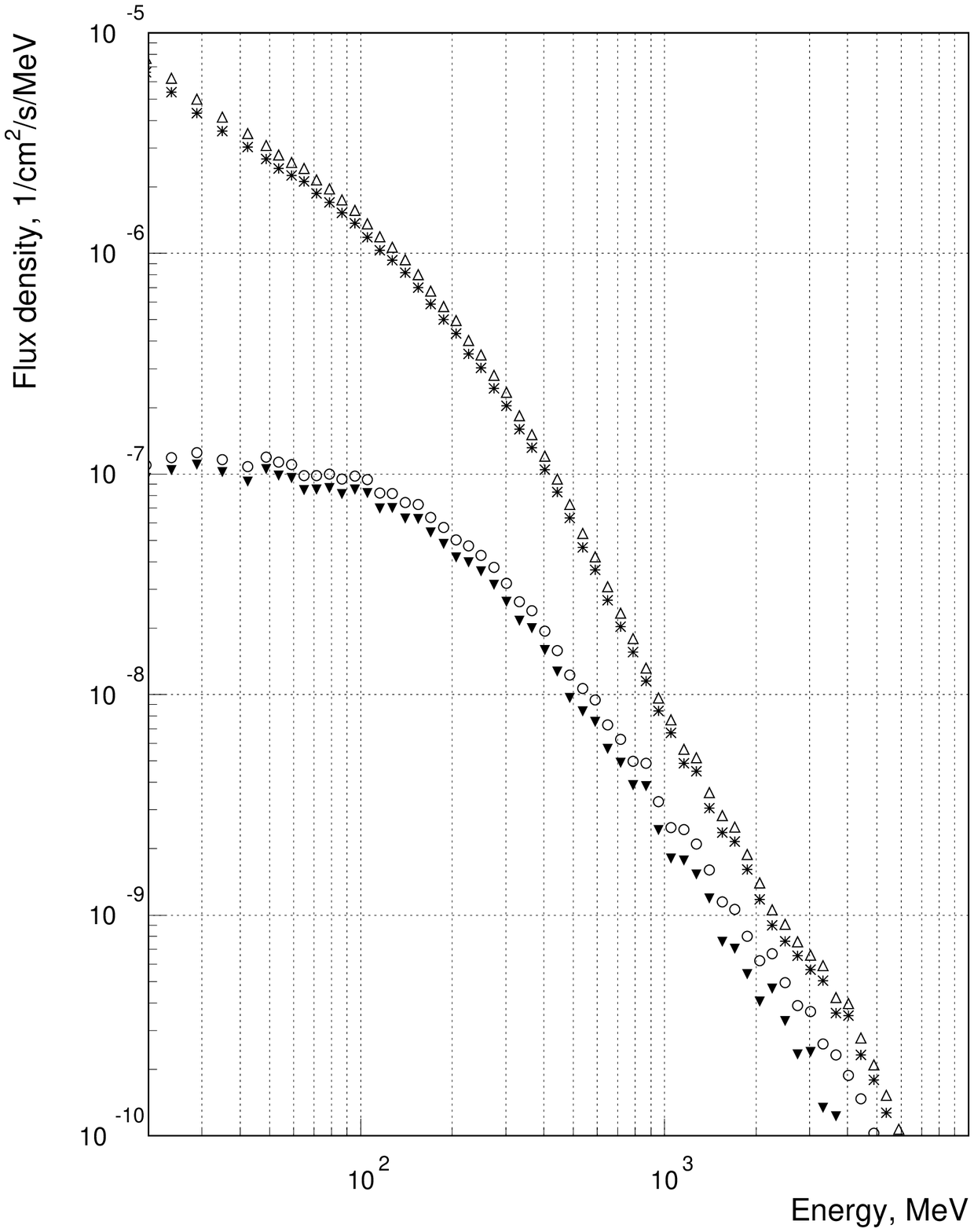,width=13cm}
\end{center}
\caption{}
\label{specinsn}   
    \end{figure*}
\newpage
\begin{figure*}[ht]
\begin{center}
\epsfig{figure=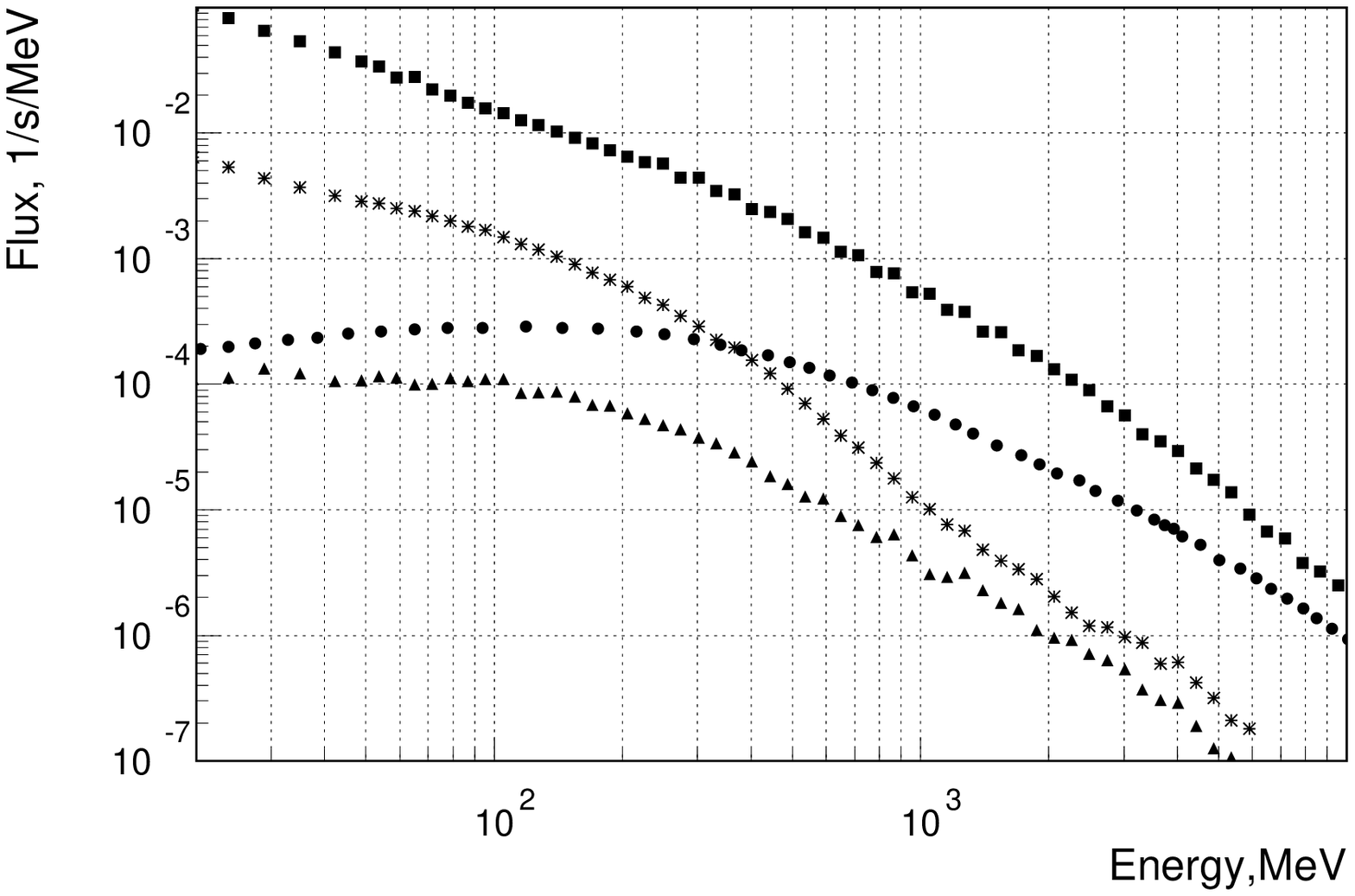,width=16cm}
\end{center}
\caption{}
\label{topatten}   
    \end{figure*}
\newpage
\begin{figure*}[ht]
\begin{center}
\epsfig{figure=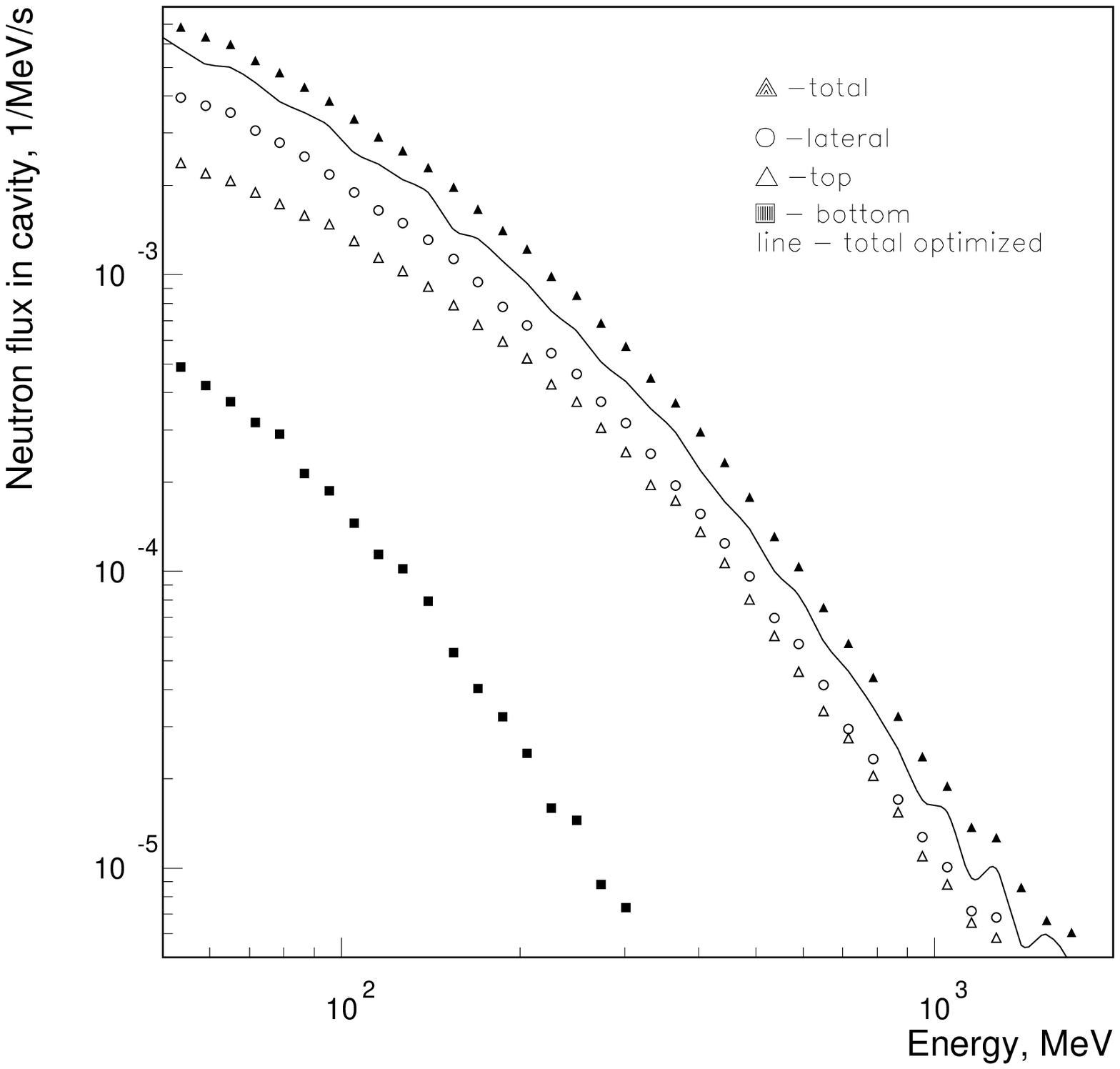,width=16cm}
\end{center}
\caption{}
\label{sides}   
    \end{figure*}

\end{document}